**Monitoring of the offset between UTC and its prediction broadcast by the GNSS**

P. Defraigne[1], E. Pinat[1], G. Petit[2], F. Meynadier[2]

[1]Observatoire Royal de Belgique, Avenue Circulaire 3, Brussels, B-1180, Belgium
[2]Time Department, Bureau International des Poids et Mesures, 92312 Sèvres Cedex, France

E-mail : p.defraigne@oma.be

**Abstract**

We present a new approach to report in the Section 4 of BIPM Circular T daily values of the offset between UTC and the predictions of UTC broadcast by the GNSS, this quantity we name $bUTC_{GNSS}$. In this approach, the determination of UTC - $bUTC_{GNSS}$ is based on data collected by several multi-GNSS stations in selected time laboratories worldwide. Test computations over a 7-month period from July 2022 to January 2023 show that the offset between UTC and $bUTC_{GNSS}$ was between 30 and 50 ns for GLONASS, between 5 and 20 ns for BeiDou, and between -5 and +5 ns for GPS and Galileo. We derive the uncertainty on the reported values, which is 4.1 ns for BeiDou and GPS, 3.7 ns for Galileo and 6.6 ns for GLONASS and show that, over the test period, the reported values of UTC–$bUTC_{GNSS}$ and the solutions obtained from each multi-GNSS station are all consistent within the 1-sigma uncertainties.

## 1. Introduction

Coordinated Universal Time (UTC) is the international reference time scale that forms the basis for the coordinated dissemination of standard frequencies and time signals. It is computed by the International Bureau of Weights and Measures (BIPM), from the clock data provided by some eighty time laboratories worldwide [1]. UTC is computed each month in deferred time so that it is only available a posteriori. For this reason, the laboratories participating to UTC maintain their own realization of UTC named UTC(k), where k is the acronym of the time laboratory. UTC is made available in the monthly Circular T [2] through the differences [UTC-UTC(k)] with associated uncertainties. UTC(k) may be disseminated in real time to the users through different means like e.g. radio transmission, or some internet protocols, providing access to a realization of UTC.

In addition, a prediction of UTC is also broadcast by the Global Navigation Satellite Systems (GNSS). Each of the GNSS indeed maintains its own internal reference time scale designed for system synchronization, called GNSSt hereafter. Then the satellites broadcast in the navigation message a prediction of the offset between this internal reference time scale and some realization of UTC, differing for each GNSS: GPS broadcasts a prediction of the offset between GPS Time (GPST) and UTC(USNO) [3], GLONASS broadcasts a prediction of the offset between GLONASS Time (GLONASST) and UTC(SU) [4], Galileo broadcasts the prediction of the difference between Galileo System Time (GST) and a realization of UTC based on 5 European UTC(k) [5] and BeiDou broadcasts the prediction of the difference between BeiDou

Time (BDT) and a realization of UTC based on UTC(NTSC) [6]. The regional systems also offer this information: the Japanese system QZSS broadcasts the prediction of the offset between QZSS Time (QZSST) and UTC(NICT) [7], and the Indian system NAVIC broadcasts the predicted difference between NAVIC Time (NAVICT) and both UTC(NPLI) and a prediction of UTC [8]. Therefore, the users can synchronize their clock on a prediction of UTC at any time, using only two broadcast parameters. The first one is the integer number of seconds between UTC and the GNSS time scales [9] which changes only at the time of new leap second insertion in UTC. The second one is the fractional part, the predicted offset modulo one second. This value is generally lower than 100 ns, as the GNSS time scales are all steered to UTC in some way. It is provided as the coefficients of a polynomial of degree zero to two, depending on the constellation and on the navigation signals, as detailed in [10].

In their navigation message, the GNSS do not broadcast directly a UTC(k), but a prediction of the offset between the GNSS reference time scale and either a UTC(k) or some other proxy of UTC and real-time users have no direct means of verification of the broadcast values. In order to provide users and GNSS providers some information on the predictions of UTC broadcast in the GNSS navigation messages, the BIPM a posteriori estimates the differences between UTC and the broadcast predictions of UTC. This information is available in Section 4 "Relations of UTC and TAI with predictions of UTC(k) disseminated by GNSS" of the Circular T [10], presently for GPS and GLONASS only, as shown in Figure 1. The reported values are computed from single frequency GNSS measurements collected by receivers calibrated from different sources. Furthermore, the associated uncertainties are purely conventional, and reported in the supplementary document as being "of the order of 10 ns for GPS and of the order of 100 ns for GLONASS". In its Recommendation 2 (2015) [11], the Consultative Committee for Time and Frequency encourages the BIPM to add similar information on new GNSS as they become operational. Thanks to recent efforts in absolute calibration of multi-GNSS receivers for hardware delays [12, 13, 14], the determination of the offset between UTC and its prediction broadcast by the GNSS is now possible for Galileo and BeiDou so that the BIPM considers adding these two constellations in the section 4 of Circular T. Only BeiDou-3 solutions will be provided as BeiDou-2 is not a constellation fully available worldwide. This paper presents the new computation strategy with an improved and consistent approach for the four constellations, as well as a complete uncertainty budget for the published values of each constellation.

```
4 - Relations of UTC and TAI with predictions of UTC(k) disseminated by GNSS.

[UTC-UTC(USNO)_GPS]  = C0',   [TAI-UTC(USNO)_GPS]  = 37 s + C0'
[UTC-UTC(SU)_GLONASS]= C1',   [TAI-UTC(SU)_GLONASS]= 37 s + C1'

For this edition of Circular T,           S0'= 0.7 ns,        S1'= 6.5 ns

 2021    0h UTC     MJD    C0'/ns   N0'   C1'/ns   N1'
         SEP 27    59484     0.5    89     17.5    89
         SEP 28    59485     0.2    90     18.3    85
         SEP 29    59486    -1.1    89     17.8    87
         SEP 30    59487    -0.8    89     19.7    84
         OCT  1    59488    -0.5    89     19.4    88
         OCT  2    59489    -0.1    90     18.4    85
```

*Fig. 1. Example of the current Section 4 in BIPM Circular T*

It can be seen in Figure 1 that the current naming of the broadcast information is UTC(USNO)_GPS and UTC(SU)_GLONASS, meaning UTC(k) as broadcast by the GNSS. As not all the GNSS constellations refer to a unique UTC(k) for their broadcast prediction of the offset between the GNSS time (GNSSt) and UTC, we use here a naming convention common to all constellations for the broadcast prediction of UTC: bUTC$_{GNSS}$, with GNSS being BDS, GAL, GLO or GPS, and BDS corresponding to BeiDou-3 as mentioned previously.

The paper is organized as follows: we first present in section 2 the new procedure to compute [UTC- bUTC$_{GNSS}$] and in section 3 the criteria used to select the receivers providing data for this computation. We then detail the uncertainty budget for each GNSS in section 4 and a 7-month test computation is presented in section 5. In section 6 we examine the differences between the [UTC- bUTC$_{GNSS}$] values obtained from dual-frequency GNSS data and those obtained with single-frequency data, and in section 7 we conclude and discuss the improvements brought by the new procedure.

## 2. Computation procedure

The difference [UTC- bUTC$_{GNSS}$] is obtained at 0:00 UTC each day by combining [UTC(k) – GNSSt] obtained from a calibrated receiver driven by a UTC(k), the correction GNSSt-bUTC$_{GNSS}$ determined from parameters broadcast in the GNSS navigation message, and [UTC – UTC(k)] from Circular T, as follows:

$$UTC - bUTC_{GNSS} = [UTC(k) - GNSSt] + [GNSSt\text{-}bUTC_{GNSS}] + [UTC - UTC(k)]_{circular\ T} \quad (1)$$

In practice, the first term [UTC(k) – GNSSt] is computed from the GNSS pseudorange measurements collected by a calibrated GNSS station connected to UTC(k), modelled using the broadcast satellite orbit and clock correction, using the standard CGGTTS [15]. The CGGTTS files report the differences between UTC(k) and the GNSSt for each visible satellite, for a set of 89 tracks per day, each with a 13-minute duration. For each track, the final value of [UTC(k) – GNSSt] at the mid-track epoch corresponds to the weighted average of the values obtained with the different satellites, using an elevation-dependent weighting. The clock solution [UTC(k) – GNSSt] is then smoothed with a 24h-window sliding average (boxcar) to mitigate the impact of the pseudorange measurement noise and multipath, and of the residual errors of the satellite orbits and clock corrections. The smoothed solution at 0:00 UTC is then retained.

The second term of equation (1) is the value at 0:00 UTC of the broadcast prediction of the offset between GNSSt and UTC. As shown in [10], due to the regular update of the navigation messages at the satellite level, different satellites from a same constellation can simultaneously broadcast different sets of parameters for the prediction of UTC, resulting in different possible values for the GNSSt-bUTC$_{GNSS}$ at a same epoch. We therefore gather the different polynomials transmitted in the preceding day and for each of them we generate the corresponding value of GNSSt-bUTC$_{GNSS}$ at 0:00 UTC of the current day. In the RINEX format version 3, only one value per constellation can be reported in the daily navigation file but it is not known how the reported message is chosen among the set of messages collected during the day. Combining the RINEX navigation files from different stations can give a first estimation of the different broadcast navigation messages. Another option is to get them directly from the raw messages available in the binary files for some receiver. Finally, with the RINEX format 4.0, different polynomials are now made available by the International GNSS Service (IGS) in the navigation files. We then retain the median of the values of GNSSt-bUTC$_{GNSS}$ at 0:00 UTC from all polynomials, to limit the impact of outliers that can be frequent for some constellations [10].

Finally, the last term of equation (1) is the value UTC – UTC(k) obtained from the first section of the BIPM Circular T. Since Circular T publishes the values of UTC – UTC(k) with one point every five days, we linearly interpolate these data to get one point per day as needed for this computation. Only very stable UTC(k) realizations are used for this purpose, so that the contribution of this interpolation to the final uncertainty is limited, as will be seen later in this paper.

The solutions UTC(k) – GNSSt can be computed from single-frequency measurements, using some external model to correct for ionospheric delays. However, an error in the ionospheric delay modeling can introduce a bias in the solution. We therefore prefer to use a linear combination of dual-frequency measurements, which removes the ionospheric delay. Only the first order ionospheric delay is removed, but higher-order effects are smaller than the pseudorange measurement noise [16]. The dual-frequency solutions are based on the following pairs of signals for each constellation: P1 (or C1) and P2 for GPS, P1 (or C1) and P2 for GLONASS, E1 and E5a for Galileo, and B1C and B2a for BeiDou.

As many users get their solution from a single frequency, generally in L1 band, we also determine in this paper the additional uncertainty associated with the possible differences between the single-frequency solutions and the dual-frequency solutions. This will be detailed in section 6.

## 3. Choice of the pivot UTC(k) to compute UTC- $bUTC_{GNSS}$

As mentioned in the previous section, the role of the pivot UTC(k) laboratory is crucial in the sense that any unquantified delay in its GNSS equipment or any instability in the time scale UTC(k) itself will add inaccuracies in the published values of UTC – $bUTC_{GNSS}$. For this reason, the choice of these pivot laboratories should be driven by calibration and UTC(k) stability. Calibration means the determination of the hardware delays in the receiving station, which are needed to get the accurate offset between a ground clock and a time scale broadcast by a satellite. Note that the satellite hardware delay is also needed, but is generally included in the broadcast satellite clock corrections. This will be discussed in more detail in section 6. The current calibration scheme in the network of time laboratories contributing to UTC is based on differential calibration using traveling equipment [17]. Since 2014, reference GNSS stations are maintained by the BIPM and by some UTC(k) laboratories, named Group 1 (G1) laboratories, selected in each Regional Metrology Organization (RMO). BIPM ensures the calibration of the G1 laboratories, using a traveling equipment, and the G1 laboratories are then responsible for the calibration of the other laboratories (called Group 2) in their RMO [18, 19]. In order to determine UTC- $bUTC_{GNSS}$ using pivot UTC(k) laboratories regularly calibrated and monitored by the BIPM, these have to be chosen among the G1 laboratories. Furthermore, the UTC(k) of the chosen G1 laboratories should be driven by a stable H-maser. Finally, a geographic distribution must be ensured. The current selection (as of 2023) includes the NIST in Boulder (USA), the LNE-SYRTE in Paris (France, UTC code OP), the NIM in Beijing (China) and the NICT in Tokyo (Japan).

It must be noted that the reference for calibration must be accurately determined for the computation of UTC – $bUTC_{GNSS}$. This is not so important for the time links contributing to

UTC, as, if the reference is biased, the same bias propagates to the full network through the relative calibration scheme described in previous paragraph. Therefore, any such bias disappears in the time links UTC(k1)-UTC(k2). However, in the current case, if a calibration bias exists, it will be present in the GNSS clock solution [UTC(k) – GNSSt] and hence contaminate also the final value of UTC – bUTC$_{GNSS}$, as seen from equation (1). For this reason, some effort has been put on absolute calibration of multi-GNSS stations in the recent years [12, 13, 14]. Absolute calibration consists in measuring the hardware delays from simulated GNSS signals, free of any satellite delay or atmospheric perturbation. It requires the availability of some dedicated equipment like an anechoic chamber for the antenna calibration, and a GNSS simulator for the receiver calibration. It is however not possible to calibrate absolutely all the stations of the time laboratories selected as pivot UTC(k). However, the BIPM reference was absolutely calibrated and then propagated to the G1 laboratories during the G1 calibration exercises. More details on the calibration uncertainties will be provided in the next section. For each constellation, the final value for UTC – bUTC$_{GNSS}$ retained for publication in the Circular T is the median of the values obtained with the four selected Group 1 stations mentioned here above, or with a subset of them in case of missing data.

## 4. Uncertainty budget

The uncertainty on the reported value for UTC – bUTC$_{GNSS}$ originates in the measurements and the computation procedure used to generate the value in deferred time. It should not be considered as the uncertainty on the broadcast value bUTC$_{GNSS}$ itself; this uncertainty is given by the GNSS provider. In other words, the uncertainty derived here and to appear in the Section 4 of Circular T cannot be used to characterize the prediction of UTC provided by the GNSS.

The contributions to the uncertainty budget on the reported UTC – bUTC$_{GNSS}$ values correspond to the uncertainties on the three terms of equation (1). For the first one, i.e. the clock solution UTC(k)-GNSSt based on pseudorange measurements, two separate components have to be considered: the receiver calibration is treated in section 4.1 and the noise of the solution in section 4.2. The second term of equation (1) is the broadcast prediction of the difference GNSSt-bUTC$_{GNSS}$. Its uncertainty is due to the lack of information on the polynomial to be used, and is taken to be the dispersion of the values from the different messages broadcast simultaneously by different satellites of the constellation, and is computed in section 4.3. Finally, the third term is UTC-UTC(k) and its uncertainty is given in the section 1 of Circular T, to which the interpolation from a five-day sampling to daily data must be added, as discussed in section 4.4. Note that this uncertainty analysis is valid for dual-frequency solutions as used to compute the first term of equation (1) UTC(k)-GNSSt. The case of single-frequency solutions will be treated in section 6.

It might be noted that, in the case that the same GNSS link is used for calculating terms 1 and 3 of equation 1, some of the systematics vanish thus lowering the global uncertainty. We expect this conjunction to be relatively rare and transient : a growing number of these UTC(k)s are or will be linked by TWSTFT, with the nature of the link possibly changing from one month to the next depending on what is most efficient for Circular T calculation. Moreover that would concern only one of the constellations out of four (also possibly different from one month to the next). As our intent is to provide a conservative value for the uncertainty, we decided not to take this possible correlation into account.

### 4.1. The uncertainties on the receiver calibration

As mentioned in section 3, the accuracy of the difference UTC – bUTC$_{GNSS}$ depends on the calibration of the GNSS station in the pivot UTC(k) laboratory, and furthermore an absolute calibration is needed for this application. Absolute calibration is currently carried out by three laboratories, CNES, ESTEC, and JPL. They all report uncertainties at a level around 1 ns 1-sigma for each modulation on each frequency, while a bit larger for GLONASS. However, in depth comparisons of these absolute calibration results show differences slightly larger as shown in [19]. This may be due to differences in the GNSS signal simulator and the correlation tools used.

The uncertainties on the calibration will be analysed here separately for each GNSS constellation. For GPS signals, the reference, maintained by the BIPM and the Group 1 laboratories has been described in detail in [19]. Even though the reference is based on a very old absolute calibration, it is shown that the differences between the hardware delays from the reference and those determined by recent absolute calibrations are always lower than 2 ns in absolute value for isolated codes. Furthermore, recent absolute calibrations carried out by different timing laboratories provide differences with the reference which have opposite signs. For this reason, the current reference for GPS, while based on a quite old absolute calibration, is considered to be consistent with more recent absolute calibrations. This reference is therefore used as well for the computation of bUTC$_{GNSS}$. For Galileo and GLONASS, the reference is based on the BIPM receiver BP21 calibrated by the ESTEC for these signals as in [12], and validated by the CNES as in [14]. For BeiDou-3 signals, a further absolute calibration was carried out by the ESTEC on another BIPM receiver (BP27). All these references were then transferred to the Group 1 laboratories through traveling equipment during G1 exercises, including the four pivots mentioned in previous section.

The absolute calibration uncertainty provided in the reports [12] and [14] is used here. It depends on the signal type, and reaches from 0.6 up to 1.0 for GPS, Galileo and BeiDou, and up to 1.5 ns for GLONASS, where only the C and P codes on the central frequencies in the L1 and L2 bands have been calibrated. Considering a common hardware delay for all GLONASS frequencies of a same frequency band is of course a big approximation. Furthermore, as we are working with dual-frequency ionosphere-free combinations, the calibration uncertainty is also increased by the linear combination. The so-obtained absolute calibration uncertainty for each constellation is presented in Table 1.

As shown in [19], the differences between the absolute calibration results obtained by different teams, can in some cases be larger than the combined uncertainties. In order to validate our proposed uncertainties, we also computed the RMS of the differences between absolute calibrations operated by different teams for the dual-frequency combinations in GPS and Galileo, using the values of the differences presented for each separate code in Tables 6 and 7 of [19]. These RMS give 2.5 ns for GPS and 1.5 ns for Galileo and are therefore quite compatible with the computed uncertainty from the absolute calibration reports. No such comparisons were available for GLONASS or BeiDou.

Finally, the uncertainty on the G1 relative calibration campaign for each receiver, obtained from the BIPM reports on G1 exercises, is added quadratically to get the final uncertainty on the calibration for the UTC – bUTC$_{GNSS}$ computed values, which gives values between 2.4 and 3.8 ns as presented in Table 1.

Table 1. uncertainties (1-σ) on the calibration for the ionosphere-free combinations (ns)

|  | BeiDou | Galileo | GLONASS | GPS |
|---|---|---|---|---|
| Absolute calibration | 2.0 | 2.0 | 3.4 | 2.3 |
| G1 calibration | 1.3 | 1.3 | 1.7 | 1.3 |
| **Combined uncertainty** | **2.4** | **2.4** | **3.8** | **2.7** |

### 4.2. GNSS clock solution noise

The clock solution GNSSt-UTC(k) obtained from the CGGTTS standard is affected by the pseudorange noise and multipath, and by the errors on the broadcast satellite orbit and clock corrections used in the pseudorange modelling. We therefore apply a smoothing (24h sliding average, boxcar) and retain the smoothed solution at 0:00 UTC of each day. The standard deviation of the differences between the raw and smoothed solutions is then used as the noise contribution to the uncertainty budget of the UTC-bUTC$_{GNSS}$. Figure 2 presents an example of smoothed solutions for each of the constellations, and for the four stations used as pivot, over the 7-month period used for validation in section 5. We observe a larger value for GLONASS, which is a consequence of the Frequency Division Multiple Access (FDMA) technique used in that constellation. The signals with different frequencies used by the different satellites have indeed different hardware delays in the receiver, with inter-frequency biases up to a ten of nanoseconds [20]. These satellite-dependent hardware delays are not taken into account in the current calibration of GNSS stations for GLONASS signals, which induces a dispersion in the clock solutions obtained by the different satellites, and hence medium-term variations in the final solution due to the variable visible part of the constellation. The standard deviations of the differences between the smoothed and raw clock solution, computed over a period of 7 months, are reported in Table 2. For each GNSS, the noise contribution to the global uncertainty budget of the reported values for UTC-bUTC$_{GNSS}$ is chosen as the maximum standard deviation among the four stations.

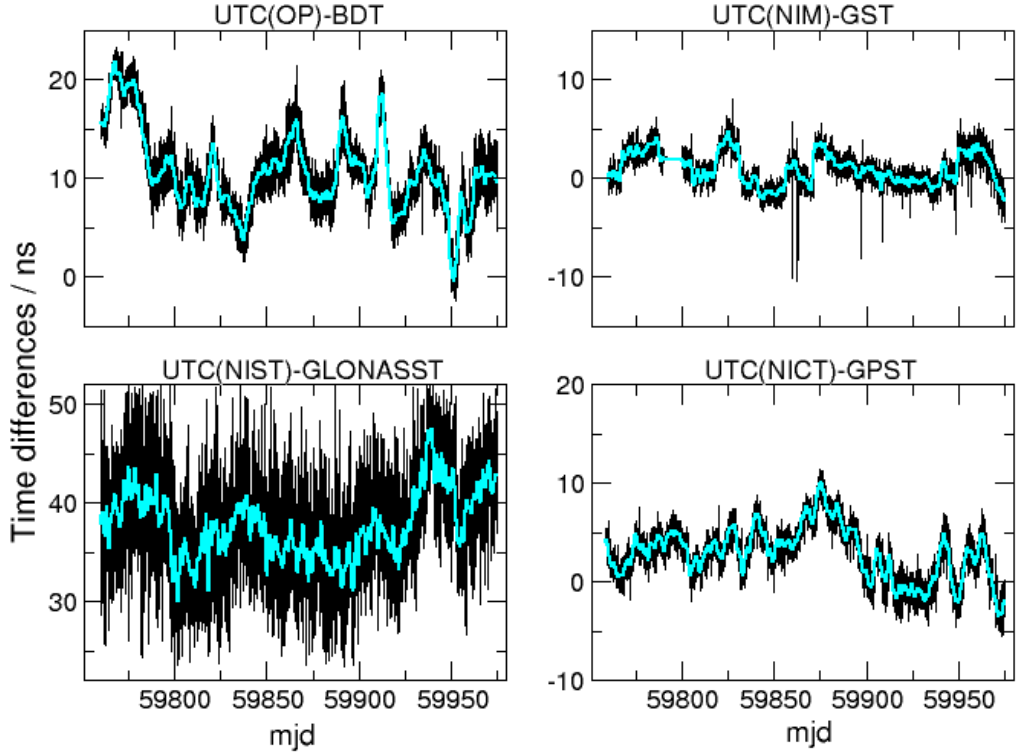

Figure 2. Raw (black) and smoothed (blue) solutions UTC(k)-GNSSt for a sample of one lab per GNSS

Table 2. rms of the differences between the smoothed and raw clock solution /ns

|  | **BeiDou** | **Galileo** | **GLONASS** | **GPS** |
|---|---|---|---|---|
| OP73 | 1.4 | 0.4 | 3.2 | 0.8 |
| NISG | 1.6 | 0.7 | 3.2 | 1.0 |
| IM15/02 | 1.8 | 0.7 | 4.6 | 0.9 |
| NC5S | 2.0 | 0.7 | 2.9 | 0.9 |
| **Maximum** | **2.0** | **0.7** | **4.6** | **1.0** |

### 4.3. Diversity of messages GNSSt–bUTC$_{GNSS}$

As explained in section 2, different satellites of a same constellation can simultaneously broadcast different messages GNSSt–bUTC$_{GNSS}$. Hence, users from all over the world can receive different messages, and can synchronize their clock on a different value of the prediction of UTC. A detailed analysis of these differences is presented in [9] for all constellations and navigation messages, and shows that for some constellations, the number of different simultaneous messages can be larger than 5. In computing the daily value (DV$_{GNSS}$) of GNSSt - bUTC$_{GNSS}$, we use the median of the values at 0:00 UTC computed with the different polynomials broadcast during the preceding day.

The uncertainty associated with this choice of DV$_{GNSS}$, considering the diversity of messages, is determined as follows. The different messages broadcast by each constellation during the last hour of the preceding day are collected from the binary files of a multi-GNSS receiver

PolaRx5TR located in Brussels, at the Royal Observatory of Belgium. The following messages have been considered: for GPS the LNAV, for Galileo the FNAV, for BeiDou B-CNAV1 and B-CNAV2 and for GLONASS the NAV message. For each message, we compute the corresponding value of $GNSSt - bUTC_{GNSS}$ at 0:00 UTC. Then, for each day of the 7-month period used for validation in section 5, we calculate the difference between $DV_{GNSS}$ and each of the values determined from the messages found in the binary file of the previous hour. Figure 3 shows the histograms of these differences. Some large differences are not visible in these plots as being larger than the X-axis range: GLONASS maximum difference is 7.4 ns, BeiDou B-CNAV1 differences present three outliers at -63.4, -30.1 and -4.4 ns, while B-CNAV2 differences with DV do not show these outliers, with a maximum difference of 2.7 ns. Put together, all these differences do not follow a normal distribution as shown in Figure 3. For this reason, the "1-sigma" uncertainty is determined as being one half of the 95th percentile value of the distribution indicated as vertical lines in Figure 3. The so-obtained uncertainties are provided in Table 3, and range between 0.1 ns for Galileo and 1.3 ns for GPS LNAV.

An alternative approach could be determining the uncertainty on the median as proposed e.g. in [21], in which case we could compute one uncertainty per day for each GNSS. However, as the data retrieved in the RINEX files can be from different moments of the day, we considered more correct to estimate one standard uncertainty per GNSS by comparing the daily value DV with the existing navigation messages around midnight, as described above.

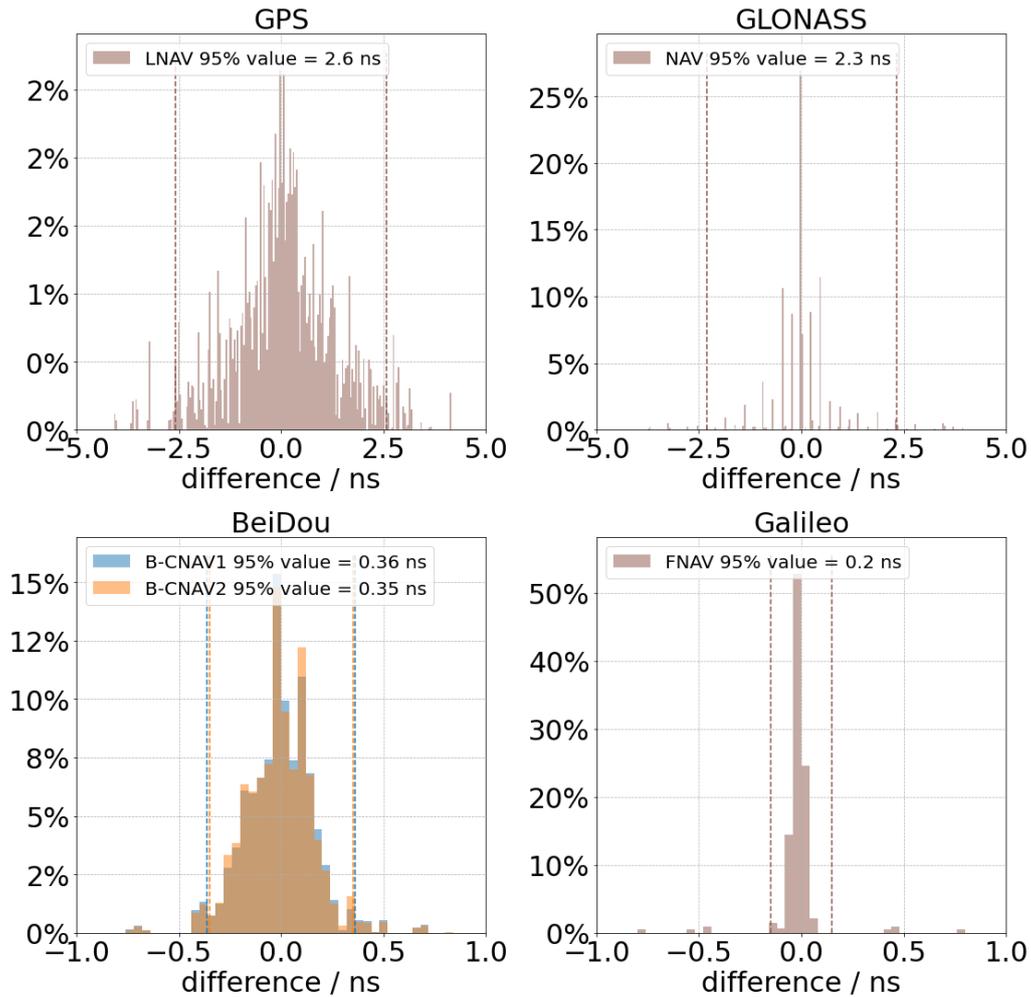

Figure 3 Histograms of the differences between the value of GNSSt–bUTC$_{GNSS}$ at 0:00 UTC used for the determination of UTC–bUTC$_{GNSS}$ and the values determined from the different messages broadcast in the last hour of the day before. The vertical lines correspond to plus and minus the "2–sigma" value determined as the 95th percentile of the distribution. Note the different the X-axis range for the different GNSS.

Table 3. Uncertainty (1-σ) due to the navigation message dispersion

|  | **BeiDou** | **Galileo** | **GLONASS** | **GPS** |
|---|---|---|---|---|
| ½ 95$^{th}$ percentile / ns | 0.2 | 0.1 | 1.2 | 1.3 |

### 4.4. Uncertainty on UTC-UTC(k)

The last component to the uncertainty budget is the one on the last term of equation (1), i.e. UTC-UTC(k). This quantity is available at a five-day interval in BIPM Circular T with a given uncertainty different for each UTC(k), and determined from the time transfer method used by each laboratory for its contribution to UTC. As this uncertainty is different for each of the UTC(k), we consider the maximum of them for the four G1 laboratories considered here. In the 7-month analysis proposed in section 5 for validation, the maximum uncertainty is 2.6 ns.

Moreover, the values of UTC-UTC(k) are interpolated linearly to get a daily sampling and report the UTC–bUTC$_{GNSS}$ on a daily basis. This can induce a difference between the physical signal UTC(k) connected to the GNSS receiver and producing the results UTC(k)– bUTC$_{GNSS}$ and the interpolated UTC(k), and an associated uncertainty when subtracting these terms to determine UTC–bUTC$_{GNSS}$ using equation (1). Considering that the UTC(k) used as pivot in equation (1) are based on a stable H-maser, with a stability not worse than 3e-15@2.5 day, we get an error on the interpolation of 0.6 ns.

### 4.5. Final uncertainty budget

Table 4 shows the final uncertainty budget as derived from the quadratic sum of the five components described in previous sub-sections These final uncertainties range from 3.7 ns for Galileo up to 6.6 ns for GLONASS. Some of components can be variable in time, as e.g. the clock noise. However our estimated uncertainty, in general based on the maximum value observed among all pivot labs, should be conservative enough to cover such variations.

Table 4. Final uncertainty budget for the computed UTC–bUTC$_{GNSS}$ / ns

|  | **BeiDou** | **Galileo** | **GLONASS** | **GPS** |
|---|---|---|---|---|
| Calibration | 2.4 | 2.4 | 3.8 | 2.7 |
| Broadcast value dispersion | 0.2 | 0.1 | 1.2 | 1.3 |
| Clock solution noise | 2.0 | 0.7 | 4.6 | 1.0 |
| UTC-UTC(k) pivot | 2.6 | 2.6 | 2.6 | 2.6 |
| Interpolation | 0.6 | 0.6 | 0.6 | 0.6 |
| **Total** | **4.1** | **3.7** | **6.6** | **4.1** |

The different components of the uncertainty budget have been estimated using the current status of the signals and calibration performances. The main contributions are from calibration, noise and UTC-UTC(k). No significant improvement or degradation is expected in the near future for the calibration. The noise for GLONASS can be improved only if a frequency-dependent calibration is made possible for GLONASS signals, or when the future generation of satellites will provide CDMA signals in two frequencies. The noise observed in the other constellation results is due to some local multipath, but also to some biases in the broadcast satellite clocks or group delays, which could improve in the future. Finally, the uncertainty on the pivot UTC-UTC(k) can also evolve with time depending on the time transfer techniques used by the laboratories. If this method is retained as the basis for future Circular T section 4, the uncertainty on UTC–bUTC$_{GNSS}$ reported there will have to be verified on a regular basis, or in case of major change of known origin in one of the components.

## 5. Validation

The technique and its associated uncertainty budget have been validated using a 7-month period, from July 2022 up to January 2023. Using the procedure described in section 2, we computed UTC–bUTC$_{GNSS}$ for the four GNSS constellations, and using the four UTC(k) mentioned previously with the following receivers: OP (OP73), NIST (NISG), NICT (NC5S) and NIM (IM15 and IM02). Two receivers are used for NIM as in the analysed period, IM15 does not provide any BeiDou-3 observations, while IM02 provides all the constellations, but starting only in October 2022. Note also that NC5S started tracking BeiDou-3 signals only on mjd 59900. All the calibration values of these receivers correspond to the G1 calibration trip 1001-2020, except for all signals of IM02 and the BeiDou signals of NC5S which are the preliminary results of the last G1 calibration in Asia, 1001-2022.

The hardware delays of the receivers used for the results presented in this paper are provided in Table 5.

Table 5. Hardware delays used in the 7-month campaign presented in this paper / ns.

|      | GPS P1 | GPS P2 | GAL E1 | GAL E5a | GLO P1/C1 | GLO P2/C2 | BDS B1C | BDS B2a | CABDLY | REFDLY |
|------|--------|--------|--------|---------|-----------|-----------|---------|---------|--------|--------|
| IM02 | -5.6   | -8.5   | -3.6   | -2.0    | -7.8      | -5.5      | -3.7    | -2.7    | 213.6  | 174.6  |
| IM15 | -27.6  | -38.3  | -27.1  | -36.4   | -28.3     | -41.6     |         |         | 212.4  | 171.3  |
| NC5S | 393.4  | 392.6  | 395.8  | 395.5   | 390.1     | 399.1     | 395.1   | 395.3   | 0      | 266.6  |
| NISG | 29.5   | 27.9   | 31.8   | 31.5    | 27.8      | 30.1      | 31.3    | 30.5    | 298.5  | 1592.2 |
| OP73 | 29.5   | 26.3   | 31.7   | 31.3    | 29.0      | 27.6      | 31.8    | 31.1    | 129.6  | 85.2   |

Figure 4 presents the results obtained for UTC–bUTC$_{GNSS}$ for each constellation over the 7-month campaign. The final solution to be kept for the section 4 of the Circular T is also depicted. As mentioned previously, it corresponds to the median of the available values (on some days, missing data did not allow to have a value for all stations). In the lower part of each plot, the differences between the median and each of the receivers is also provided. These differences are fully in line with the 1-σ uncertainties computed in previous section. The drift observed in NC5S results is associated to a seasonal variation of the station hardware delay with an annual periodicity and a magnitude of 2.5 ns peak to peak in both Galileo and GPS, which could be observed on a dedicated analysis over the years 2021 and 2022 and could be due to some temperature sensitivity.

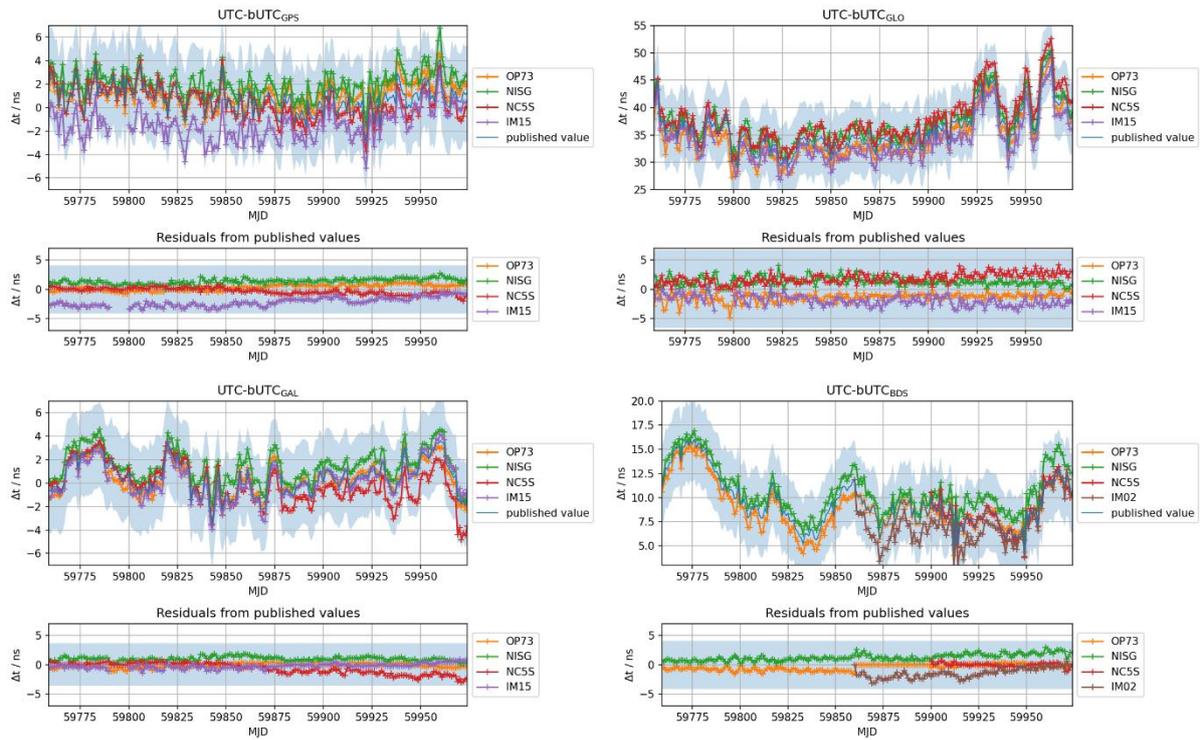

Figure 4. UTC−bUTC$_{GNSS}$ (and its uncertainty in light blue) as computed from the four G1 stations for each constellation, and differences with respect to the median used as final solution.

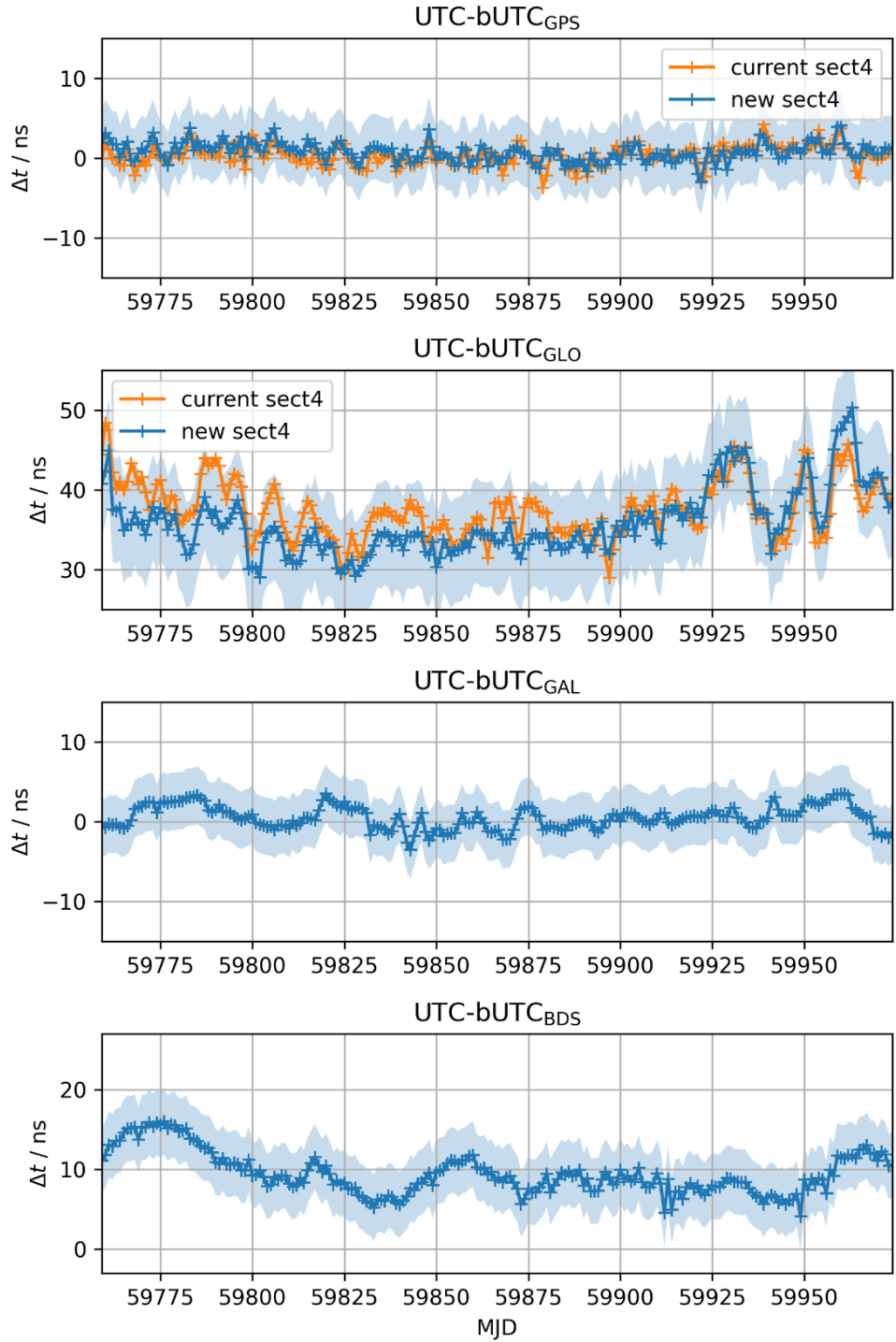

Figure 5. UTC−bUTC$_{GNSS}$ final values obtained as the median of solution obtained with the available stations among the four G1 laboratories from July 2022 to January 2023. Blue-shaded areas above and below the curve correspond to the total uncertainty budget in Table 4. For GPS and GLONASS, the "current sect4" orange curve shows the results that have been published, using the current method, in Circular T section 4.

Figure 5 presents the final results of UTC−bUTC$_{GNSS}$ for the period used for validation, i.e. from July 2022 to January 2023. In theory, i.e. neglecting all other sources of uncertainty, a user synchronizing his/her clock using the prediction of UTC broadcast by GNSS during that period had therefore an offset to UTC between 30 and 50 ns if using GLONASS, between 5 and 20 ns if using BeiDou-3, and between -5 and +5 ns if using GPS or Galileo. The uncertainties associated to these offsets are 4.1 ns for BeiDouS and GPS, 3.7 ns for Galileo and 6.6 ns for GLONASS, as detailed in the previous section. Figure 5 also shows a very good consistency with the results currently reported in the Circular T for GPS and GLONASS, based on single-frequency receivers located in OP for GPS, and AOS for GLONASS. The difference with the current estimation and the new procedure is within the uncertainty of the newly computed values.

## 6. Difference between single- and dual-frequency solutions

All the clock solutions presented here above are based on dual-frequency combination to get rid of the ionospheric delays. However, a high number of users are still determining their time synchronization from single-frequency receivers, using classically the signal on the highest frequency band L1, for which the ionospheric error is the lowest. We therefore consider in this section the differences that can exist between single- and dual-frequency solutions. It is clear that, whatever the signal or the combination used, the clock synchronization computed from GNSS measurements should be equal if the equipment has been fully calibrated. However, as shown in Figure 6, differences of a tens of ns can exist between the clock solutions obtained from single or dual-frequency solutions. These differences can be explained by three different factors:

- the receiver calibration errors: in single frequency, only the calibration of one modulation on the L1 frequency is used. Any calibration error on this signal or on the second signal used for the ionosphere-free combination can lead to a difference between the single- and dual-frequency solutions.

- the errors on the broadcast satellite group delays. The satellite clock errors broadcast by the satellites contain the satellite hardware delays of a given signal or signal combination. For GPS, it is the dual-frequency combination of P1 and P2 [3], for GLONASS, it corresponds to the L1 frequency band [4], for Galileo it corresponds to the ionosphere-free combination of E1 and E5a (for the FNAV message used here) [5]. For BeiDou, the broadcast satellite clock errors correspond to the B3I signal [6]. The broadcast group delays are the inter-signal satellite hardware delays, and allow the user to get accurate measurements on any signal or signal combination, while using the broadcast satellite clock errors based on another signal or combination. The way the satellite group delays are computed by the system operators is not known, and any error on them will lead to differences between the receiver clock synchronization based on a single frequency signal or a dual-frequency combination.

- the errors on the ionospheric delay correction applied to single-frequency measurements. In the current study we used global ionosphere maps provided by the International GNSS Service [22]. These maps have been generated from the GNSS measurements of the IGS network and offer a spatial resolution of longitude 5° and latitude 2.5°, and a time resolution of 2h. They can therefore only represent the major structures of the ionosphere and not the local perturbations. The accuracy of the IGS maps is at the level of 1 to 3 units of Total Electron Content [23] depending on the solar activity, which, converted in associated delay on the GNSS signal in the L1 frequency band, gives up to 1.7 ns for a satellite at the zenith, or 3.4 ns for a satellite at 30° elevation. Of course, this corresponds to a global average of the maps over the world, while the accuracy can also depend on the location due to the non-homogeneous distribution of the IGS stations on the globe.

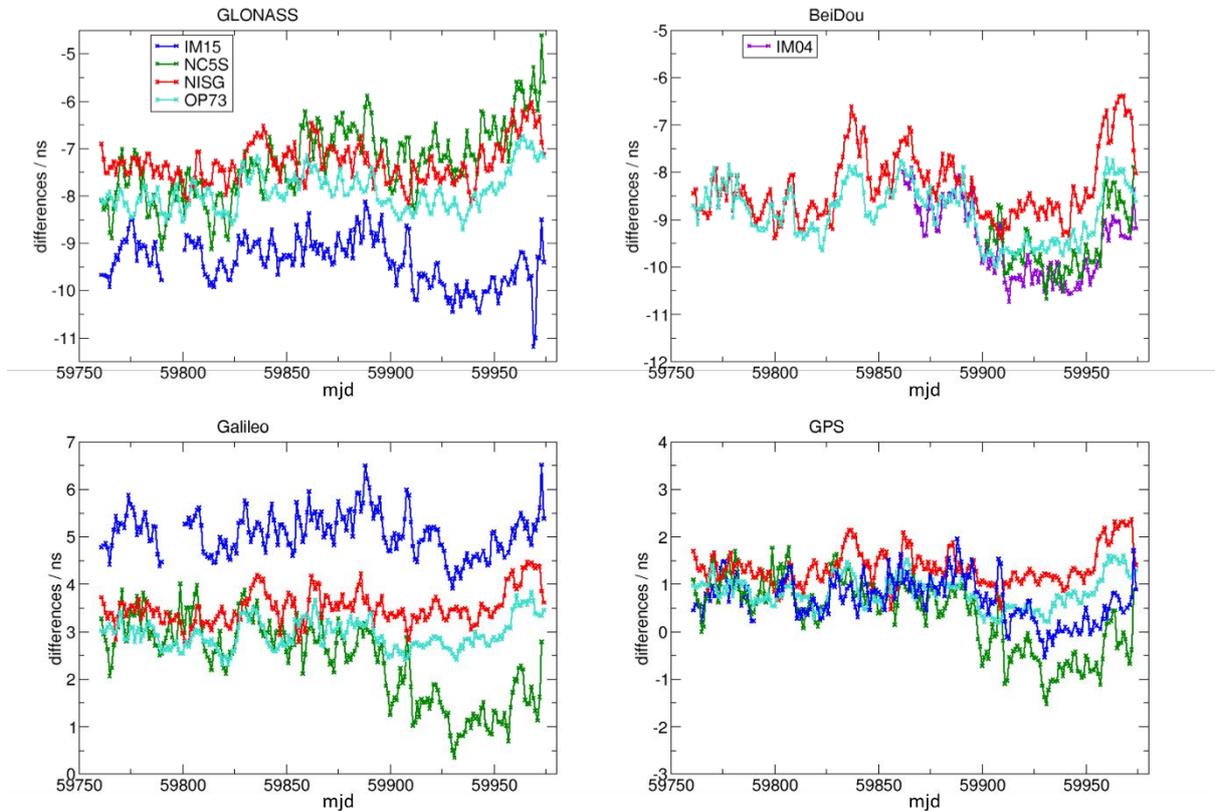

Figure 6. Differences between single-frequency and dual-frequency solutions for UTC(k) $-bUTC_{GNSS}$

The ionospheric correction in single-frequency solutions is the major cause of the variations observed in Figure 6, as receiver hardware delays can be considered as constant in view of the repeatability of calibration exercises [19] and the satellite group delays also should be quite constant. Furthermore, for a same station, we observe similar variations in the four constellations for the differences between single-frequency and dual-frequency clock solutions depicted in Figure 6. This is visible for all constellations on the short term (a few days), while the longer term variations are similar for GPS, Galileo and BeiDou, but not for GLONASS for which some errors in the variable satellite group delays can be more important.

## 7. Discussion and conclusion

We have presented and validated a new approach for determining the offset between UTC and the predictions of UTC broadcast by the GNSS in their navigation messages, to be reported in the Section 4 of BIPM Circular T. This approach makes use of a number of multi-GNSS calibrated stations (presently four) located in some pivot UTC(k) laboratories. These are chosen for their participation in regular BIPM calibrations, hence must be Group 1 laboratories, with a stable UTC(k) time scale, and distributed geographically over the world. The laboratories used to start this new approach are NIST, OP, NIM and NICT. The values reported in the

Circular T correspond to the median of the solutions computed with the four stations (or less if one of them is not available for a limited period of time). The uncertainty on the reported values of the difference UTC–bUTC$_{GNSS}$ has been determined considering the calibration, the noise of the GNSS clock solution, the diversity of simultaneous broadcast messages providing GNSSt–bUTC$_{GNSS}$ and the uncertainty on UTC–UTC(k). All this confers an uncertainty of 4.1 ns for BeiDou and GPS, 3.7 ns for Galileo and 6.6 ns for GLONASS. Of course, this has to be understood as the uncertainty on the computed value for UTC–bUTC$_{GNSS}$, and not on the bUTC$_{GNSS}$ itself. As an example, from July 2022 to January 2023, the offset between UTC and bUTC$_{GNSS}$ was between 30 and 50 ns for GLONASS, between 5 and 20 ns for BeiDou, and between −5 and +5 ns for GPS and Galileo.

The approach and the uncertainties were validated using data covering a 7-month period. The differences between the UTC–bUTC$_{GNSS}$ reported values (i.e. the median) and the solutions obtained from each G1 station separately are all lower than the 1-sigma uncertainty. These uncertainties will be monitored on a regular basis, and could be modified in case of major change of known origin in one of the components of the uncertainty budget.

The provided uncertainties correspond to the values computed from a dual-frequency ionosphere-free combination. Differences up to 12 ns have been shown with respect to the results of a single-frequency use of the GNSS signals. These are due to both satellite and station hardware delay uncertainties, and the uncertainties on the ionospheric delay corrections in the single-frequency approach. In the case of GPS and GLONASS, which are already part of the Circular T section 4, the difference between the published value (obtained from single receivers, with a different process) and the value obtained with the method described in this paper stays within the uncertainties computed in this paper.

The method proposed in this paper provides a reliable a posteriori estimation of the stability and accuracy of the bUTC$_{GNSS}$ broadcast values and should help determine to which level the predictions of UTC from the different GNSS can be considered equivalent. A good level of equivalence would be a significant step towards providing time interoperability for multi-GNSS analyses in positioning navigation and timing only using presently broadcast information. In this respect, our analysis also provides some insight on the achievable performance of time interoperability through bUTC$_{GNSS}$.

However, we would like to stress that, even if implemented as suggested in this paper, the section 4 of Circular T should be used for information only. For a demonstration of metrological traceability as proposed in [24], the approach described in this paper and its uncertainties would have to be formally approved under the international conventions. The BIPM's insight and verification possibilities on section 4 data is not on a par (and will not be in the foreseeable future) with what is done for section 1 (UTC-UTC(k) values, basis of the Key Comparison "CCTF-K001.UTC"), which involves constant dialogue and technical information sharing with the operators of UTC(k) timescales.

Acknowledgements :

The authors wish to thank Raphaël Valceschini, who developed an early version of the software implementing this method, while under contract at ESA, in collaboration with GP. We are also very grateful to the IGS (International GNSS Service) for collecting, compiling and making available the multi-GNSS navigation data, as well as ionospheric products, that enable us to derive part of the data presented here. The rest of the data is being provided by LNE-SYRTE – Observatoire de Paris (France), NICT (Japan), NIST (USA) and NIM (China): may their respective teams be thanked here for their continuing support and commitment.
**REF**

1. Panfilo G and Arias F 2019 The Coordinated Universal Time (UTC) *Metrologia* **56** 042001
2. BIPM Circular T [Online] *Available at*: https://www.bipm.org/en/time-ftp/circular-t.
3. Flores A 2020 Interface Specification Document IS-GPS-200L, *Available at:* https://www.gps.gov/technical/icwg/
4. GLONASS Interface Control Document Navigational radio signal in bands L1, L2 (Version 5.1) 2008, *Available at:* http://www.glonass-svoevp.ru/index.php?option=com_content&view=article&id=146&Itemid=305&lang=en.
5. Galileo Open Service Service Definition Document (OS SDD), Issue 1.1 , 2019, *Available at*: https://www.gsc-europa.eu/electronic-library/programme-reference-documents.
6. China Satellite Navigation Office 2019 BeiDou Navigation Satellite System Signal In Space Interface Control Document Open Service Signal B1I (Version 3.0) *Available at:* http://en.beidou.gov.cn/SYSTEMS/ICD/.
7. Quasi-Zenith Satellite System Interface Specification Satellite Positioning, Navigation and Timing Service (IS-QZSS-PNT-004) 2021 *Available at:* https://qzss.go.jp/en/technical/ps-is-qzss/ps-is-qzss.html.
8. Bhardwajan A, Dakkumalla A, Arora A, Ganesh T and Sen Gupta A 2021 Navigation with Indian Constellation and its Applications in Metrology *MAPAN* **36**, 227–236
9. Lewandowski W and Arias F 2011 GNSS times and UTC *Metrologia*, **48** S219–S224
10. Pinat E and Defraigne P 2022 In-depth analysis of UTC information broadcast in GNSS navigation messages *GPS Solut.* **26** 105
11. Consultative Committee for Time and Frequency (CCTF) 2015 Report of the 20th meeting to the International Committee for Weights and Measures, *Available at:* https://www.bipm.org/documents/20126/30132374/cc-publication-ID-371/b3f5e382-5e0d-fd49-5622-f0823261662f
12. Garbin E, Defraigne P, Krystek P, Piriz R, Bertrand B and Waller P 2019 Absolute calibration of GNSS timing stations and its applicability to real signals *Metrologia* **56** 015010
13. Waller P, Valceschini R, Delporte J and Valat D 2019 Cross-calibrations of multi-GNSS receiver chains Proc. of Int. Frequency Control Symp. & the European Frequency and Time Forum (https://doi.org/10.1109/FCS.2019.8856134)
14. Valat D and Delporte J 2020 Absolute calibration of timing receiver chains at the nanosecond uncertainty level for GNSS time scales monitoring *Metrologia* 57 025019
15. Defraigne P and Petit G 2015 CGTTS-V2E: an upgraded standard for GNSS time transfer *Metrologia* **52-6** G1

16. Pireaux S, Defraigne P, Wauters L, Bergeot N, Baire Q and Bruyninx C 2010 Higher-order ionospheric effects in GPS time and frequency transfer *GPS solutions* **14(3)** 267-277
17. Esteban H, Palacio J, Galindo F J, Feldmann T, Bauch A and Piester D 2010 Improved GPS-based time link calibration involving ROA and PTB *IEEE Trans. Ultrason. Ferroelectr. Freq. Control* **57** 714–20
18. BIPM guidelines for GNSS calibration, *Available at:* https://webtai.bipm.org/ftp/pub/tai/publication/gnss-calibration/guidelines/bipmcalibration_guidelines_v40.pdf
19. Petit G and Defraigne P 2023 Calibration of GNSS stations for UTC *Metrologia* **60** 025009
20. Harmegnies A, Defraigne P and Petit G 2013 Combining GPS and GLONASS in All in View for time transfer *Metrologia* **50(3)** 277–287
21. Ratel, Guy, 2006, Median and weighted median as estimators for the key comparison reference value (KCRV), *Metrologia*, **43**, 4, p. S244-S248
22. Hernández-Pajares M, Juan JM, Sanz J, Orus R, Garcia-Rigo A, Feltens J, Komjathy A, Schaer SC and Krankowski A 2009 The IGS VTEC maps: a reliable source of ionospheric information since 1998. *J Geodesy* **83(3–4)** 263–275
23. Wielgosz P, Milanowska B, Krypiak-Gregorczyk A. *et al.* Validation of GNSS-derived global ionosphere maps for different solar activity levels: case studies for years 2014 and 2018. *GPS Solut.* **25** 103
24. Defraigne P, Achkar J, Coleman M, Gertsvolf M, Ichikawa R, Levine J, Uhrich P, Whibberley P, Wouters M and Bauch A 2022 Achieving traceability to UTC through GNSS measurements, *Metrologia* **59** 064001